\documentstyle[prb,aps,twocolumn,epsfig]{revtex}
\begin{document}

\title{AC-conductance 
of a quantum wire with electron-electron interaction}
\author{G. Cuniberti$^{1}$, M. Sassetti$^{2}$ and B. Kramer\\
I. Institut f\"ur Theoretische Physik, Universit\"at Hamburg\\
Jungiusstra\ss{}e 9, D-20355 Hamburg, Germany}
%%%%%%%%%%%%%
\author{
{\small (Received 30 July 1997, to appear in Phys. Rev. B)}
\vspace{3mm}
}
\author{%
\parbox{14cm}{\parindent4mm\baselineskip11pt%
%\begin{abstract}
{\small
    The complex ac-response of a quasi-one dimensional electron system in the
  one-band approximation with an interaction potential of finite range is
  investigated. It is shown that linear response is exact for this model. The
  influence of the screening of the electric field is discussed. The complex
  absorptive conductance is analyzed in terms of resistive, capacitive and
  inductive behaviors.
}
%\end{abstract}
\vspace{4mm}
}
}
\author{
\parbox{14cm}
}
\maketitle
%%%%%%%%%%%%%

%\begin{abstract}
%\baselineskip16pt
%\end{abstract}
%\baselineskip22pt

\newpage
\section{Introduction}

Experimental and theoretical investigations of the ac-transport in
nano-structures are of profound scientific interest since they provide
insight into the behavior of (open) quantum systems in non-equilibrium
that are externally controllable within wide ranges of parameters. In
addition, possible applications of nano-structures in future electronic
devices, which will have to operate at very high frequencies, require
detailed knowledge of their frequency and time dependent transport
behavior.

Electron transport in nano-structures is very strongly influenced by charging
effects. Most striking is the Coulomb-blockade \cite{al86} of the dc-current
through tiny tunnel junctions when the bias voltage and the temperature are
smaller than the ``charging energy'', $E_{C}=e^{2}/2C$ ($e$ elementary charge,
$C$ ``capacitance of the tunnel junction''). The use of a capacitance has been
justified by observing that its values determined from $E_{C}$ are consistent
with those obtained from the geometry of the junction \cite{fd87,getal89}; $C$
was found to be of the order of $10^{-15}$F and much smaller for metallic
junctions \cite{in92}.  Interactions also dominate transport through islands
between two tunnel contacts in series in a semiconductor quantum wire. The
{\em linear} conductance shows pronounced peaks \cite{meirav90} if the
external chemical potential coincides with the difference between the ground
state energies of $N+1$ and $N$ electrons in the island. In the charging
model, these energies are again given in terms of a ``capacitance $C$ '',
$E(N)=N^{2} e^{2}/2C$, $C\approx 10^{-15}$F. One can ask, how small a
capacitance can be without being influenced by quantum effects. The
limitations of the charging model become obvious in the non-linear transport
properties: fine-structure in the current-voltage characteristic is related to
the quantum properties of the interacting electrons
\cite{jetal92,wetal93,weinprl,weinmann96}.

In recent years, frequency dependent electrical response of systems with
reduced dimensionality became the subject of activities. These techniques are
of particular interest since no current and voltage probes have to be attached
to the sample. The current response to microwave and far-infrared radiation on
the transport through semiconductor microstructures has been studied
\cite{sollner83,chitta92,cai90,wagner}. It has been found that the absorption
of microwaves leads to a characteristic broadening of the conductance peaks of
semiconductor quantum dots in the Coulomb-blockade region
\cite{kouwen94,blick95}. Infrared absorption \cite{sm89,h91,mhp92} of quantum
dots and wires provided mainly information on the parts of the excitation
spectrum that are only weakly influenced by the interaction due to Kohn's
theorem. However, Raman scattering from quantum wires and dots showed
signatures of the dispersion of the collective excitations
\cite{goni91,setal94,schetal96}. The absorption of microwaves in an ensemble
of metallic grains has been investigated experimentally and theoretically in
many papers \cite{cpt84,zsta96}.

Absorption and scattering of electromagnetic radiation are only one
possibility to measure ac-transport properties without applying voltage and
current probes that may disturb the system's properties severely. More
recently, other highly sophisticated, non-invasive techniques for determining
ac-conductances have been pioneered \cite{rb96}. Coupling a system of about
$10^{5}$ mesoscopic rings to a highly sensitive superconducting micro-resonator
the perturbation of the resonance frequencies and quality factors has been
used to determine the real and imaginary parts of an ac-conductance. Here, the
fundamental question arises, what the differences between the
``conductances'' as determined by different methods are \cite{i86}.

Theoretical approaches have been developed, ranging from semi-classical rate
equation approximations to fully quantum mechanical attempts. The linear
theory of ac-quantum transport has been restricted to non-interacting systems
\cite{velicky89,liu91,brandes93,fu93,wjm93}. How to define quantum
capacitances \cite{brandes93} and inductances \cite{liu91,fu93} was addressed.
The ac-transport through mesoscopic structures in the presence of Coulomb
interaction was considered by using a self-consistent mean field method
\cite{buettiker93,bc96}. This approach strongly relies on the presence of
``reservoirs'', ``contacts'' and ``electro-chemical potentials'' which are not
necessary ingredients of high frequency experiments such as the absorption of
microwaves.

Photo-induced transport through a tunnel barrier \cite{tien63} and tunneling
through semiconductor double-barrier structures \cite{inarrea94} has been
considered. Charging effects in small semiconductor quantum dots in the
presence of time-varying fields were treated \cite{wjm93,bruder94}. The
influence of high frequency electric fields on the linear and non-linear
transport through a quantum dot with infinitely strong Coulomb repulsion
\cite{hettler95} was studied.  Photon-assisted tunneling through a double
quantum dot has been investigated by using the Keldish technique \cite{sw96}.
The photo-induced transport through a single tunnel barrier in a 1D
interacting electron system has been investigated \cite{sasswk96}. In most of
the latter works, quantum effects of the interaction have been treated only
approximatively.

In view of the importance of the interaction for the ac-properties of
nano-structures, Luttinger systems are of great interest.  Here, the
interaction can be taken into account exactly. The conductance of a tunnel
barrier in a Luttinger liquid with zero-range interaction has been shown
\cite{kane92,yue94} to scale with the frequency as $\omega^{2/g-2}$ ($g$
interaction parameter, see below). At
$\omega =0$ repulsive electron-electron interaction suppresses tunneling
completely, even for a vanishingly small potential barrier. However, it
has been also shown that for $\omega \to 0$ there is a {\em displacement
contribution} to the current which can dominate the transport for very strong
repulsive interaction and very high barrier \cite{guinea94}.

The driving voltage has been {\em assumed ad hoc} in most of these works.
Since the driving electric field is determined by the interaction between the
electrons \cite{sassetti96,keller96}, the current can be expected to depend on
how charges induced by an external electric field are rearranged by the
interaction. Even in the limit of dc-transport through a Luttinger system this
has been argued to be so \cite{kawabata,finkelstein}.  The dependence of the
ac-properties of a tunnel barrier in Luttinger system on the shape of the
electric driving field has been investigated \cite{sassetti96}. It has been
found that the current depends only on the integral over the driving electric
field -- the external voltage -- only in the dc-limit, even in the non-linear
regime. One of the side results was to confirm that linear response is exact
for the ideal Luttinger system \cite{sassetti96,ponomarenko,fn96}. The
ac-properties of the Luttinger liquid with spatially varying interaction
strength were also studied \cite{ponomarenko,fn96,maslov}.

In this paper, we concentrate on the ac-transport properties of electrons
described by the Luttinger model with an interaction
of finite range. The model is exactly solvable. Since its current response
can be determined without approximations, answers to fundamental
questions can be found, such as (i) to identify the specific signatures
of the electron-electron interaction in ac-transport, (ii) to
understand the influence of the properties of the driving electric
field, how to define conductances, and (iv) to understand ac-transport
in terms of resistive, capacitive and inductive behaviors. The latter
yield quantum analogs of impedances that are common in classical
electrodynamics. Such parameters are also often used for describing the
transport in nanostructures. Therefore, a quantum approach towards
their definition is highly desirable. However, in the quantum
regime, they depend not only on the interaction, but also on the
frequency and the shape of the applied electric field. Together with
their microscopic definitions, it is thus important to determine the
range of parameters for their validity.

The interaction potential is obtained by using a 3D screened Coulomb potential
(screening length $\alpha ^{-1}$) and projecting to a quasi-1D quantum wire of
a finite width $d$. When $\alpha ^{-1}\ll d$ we recover the Luttinger liquid
with zero-range interaction. On the other hand, when $\alpha ^{-1}\gg d$ we
obtain the 1D-analog of the Coulomb interaction. The excitation spectrum of
this model shows an inflection point at a frequency $\omega _{\rm p}$ that
increases monotonically with the interaction strength, and with the inverse of
a characteristic length associated with the interaction \cite{cuni96}.

By using linear response theory we obtain the complex, frequency-dependent
non-local conductivity given by the current-current correlation function. It
contains the dispersion relation of the elementary excitations, and turns out
to be independent of the temperature within the model. There are
no non-linear contributions to the current in this model.

The conductivity describes the current response of the system with respect to
an electric field $E(x',\omega )$. In an experiment, either the current as a
function of an {\em external voltage} (in the dc-limit), or the absorption of
electromagnetic energy at frequency $\omega $ from an external field is
determined. In both cases, not too much is known about the internal field.
Therefore, it is reasonable to search for quantities that do not depend on the
spatial form of the field. One possibility is ~\cite{fisherandlee} to define
the conductance $\Gamma _{1}$ by using the absorbed power. One finds that, in
the dc-limit, the result is indeed independent of the shape of the field.
However, in ac-transport, the shape of the square of the Fourier transform of
the electric field appears as a multiplicative factor.  The remaining factor
is the density of collective excitations of the interacting electrons. It has
a resonance at $\omega _{\rm p}$. For non-interacting electrons $dq/d\omega
=v_{\rm F}^{-1}$ and $\Gamma _{1}(\omega )$ reveals merely the structure of
the electric field.

That the ac-conductance depends on the shape of the driving electric field
leads to the question what the nature of this field is in an interacting
system. We show that for the conductivity of an ideal Luttinger liquid, it is
the {\em external electric} field which has to be used since linear response
is exact. Furthermore, we will demonstrate that this is also true for the
absorptive conductance.

Having determined the absorptive part of the conductance, $\Gamma
_{1}(\omega )$, the reactive part $\Gamma _{2} (\omega )$ may be
obtained by Kramers-Kronig transformation. The complex conductance
$\Gamma (\omega )=\Gamma _{1}(\omega )+{\rm i} \Gamma _{2}(\omega )$
relates the average current with the voltage. The current as a function
of time consists of two parts. One, $\propto \Gamma _{1}$, is in phase
with the driving field. The second, $\propto \Gamma _{2}$, is
phase shifted by $\pm\pi /2$.

When $\omega $ is small, we can expand
\begin{equation}
\Gamma _{1}(\omega )=g\frac{e^{2}}{h} + \gamma _{1}\omega ^{2}
        +\mbox{O}(\omega ^{4}).
\label{gamma1low}
\end{equation}
The first term corresponds to
the quantized contact conductance \cite{landauer70}
$R_{\rm K}^{-1}=e^{2}/h$. It is here
renormalized by the interaction parameter 
\cite{kane92} $g$. The term $\propto \omega ^{2}$ indicates whether the
current is capacitive ($\gamma _{1}>0$) or inductive ($\gamma_{1}<0$).
For $\Gamma _{2}$ we find
\begin{equation}
\Gamma _{2}(\omega )=\omega \gamma _{2} + {\rm O}(\omega ^{3})\,. 
\label{gamma2low}
\end{equation}
This quantity also indicates if the system behaves capacitively and
inductively, $\gamma _{2}<0$ and  $\gamma _{2}>0$, respectively.
However, in the latter case, $\gamma _{1}$ can still be positive
indicating capacitive behavior of the real part of the conductance. 

For frequencies close to the resonance, the Kramers-Kronig
transformation gives
\begin{equation}
\Gamma _{2}(\omega )\approx \gamma _{\rm m}(\omega -\omega _{\rm m}^{*}),
\end{equation}
with $\gamma _{\rm m}>0$ when $\gamma _{2} <0$ and $\omega _{\rm
m}^{*}\approx \omega _{\rm m}$ (position of
maximum of $\Gamma _{1}(\omega )$). The quantity $\gamma _{\rm m}$ indicates
capacitive and/or inductive behavior close to
the resonance frequency ($\approx \omega _{\rm p}$).

The height or, equivalently, the width of the resonance in $\Gamma _{1}$
defines also a resistance, $R$. In contrast to the contact resistance, it is
truly ``dissipative'' and related to the pair excitations of the Luttinger
liquid. It is also contained in $\gamma _{1}$, though its numerical value for
small $\omega $ is different from the one near the resonance.  Generally, we
find that it is only possible to define resistances, capacitances and
inductances in certain limited parameter regions \cite{cuni96}.

In order to observe capacitive behavior, the interaction between the
electrons should be {\em sufficiently long range}. This is consistent
with the results of a different approach in which Coulomb blockade
behavior at a tunnel barrier between two Luttinger liquids has been
discussed \cite{sassk96}. Also there, non-zero range of the
interaction is necessary for capacitive behavior.

In the next section, we describe briefly the model and the dispersion
relations for various interaction potentials. We calculate the ac-conductance
and study external versus internal driving fields in the section 3. Section 4
contains the identification of quantum impedances.  Section 5 contains the
discussion of the results.

\section{Luttinger liquid with long-range interaction}

\subsection{Outline of the model}

The Luttinger liquid is a model for the low-energy excitations of a 1D
interacting electron gas \cite{luttinger63,lp74,s79,haldane81}. Its excitation
spectrum can be calculated analytically. Also many of the thermodynamical and
the transport properties, as the linear conductivity, can be determined, even
in the presence of perturbing potentials. The main assumption is the
linearization of the free-electron dispersion relation near the Fermi level.
The starting point is the Hamiltonian for interacting Fermions with, say,
periodic boundary conditions \cite{fabrizio}
\begin{eqnarray}
H&=&\hbar v_{\rm F}\sum_{k,s=\pm}(sk-k_{\rm F})(c^{\dagger}_{ks}c_{ks} -
        \langle c^{\dagger}_{ks}c_{ks}\rangle_{0}) +\nonumber\\
&&+ \sum_{k_{1},s_{1}\cdots k_{4},s_{4}}
        V_{k_{1},s_{1}\cdots k_{4},s_{4}}
        c^{\dagger}_{k_{1}s_{1}}c^{\dagger}_{k_{2}s_{2}}
        c_{k_{3}s_{3}}c_{k_{4}s_{4}}.
\label{fermion}
\end{eqnarray}
Here, $c^{\dagger}_{ks}$, $c_{ks}$ are the creation and annihilation
operators for Fermions in the states $|ks\rangle$ of wave number
$k=2\pi n/L$ ($n=0,\pm1,\pm2,\cdots$) in the branches $s=\pm$, $k_{\rm F}$
the Fermi wave number, $V$ the Fourier-transform of the 
electron-electron interaction, and $\langle \cdots \rangle _{0}$
denotes an average in the ground state.

Formally, the Fermion Hamiltonian can be transformed into a Bosonized
form. For spinless particles with an interaction that depends only on
the distance between the particles, $V(|x-x'|)$, and taking into account
only forward scattering, one obtains a bilinear form in the
Boson operators which can be diagonalized by a Bogolubov transformation
\cite{haldane81}. The result is
\begin{equation}
H=\sum_{q}\hbar\omega (q)b^{\dagger}_{q}b_{q}\,.
\label{diagonal}
\end{equation}
The spectrum of the pair excitations corresponding to the Bosonic creation
and annihilation operators $b^{\dagger}_{q}$, $b_{q}$ is given by the
Fourier transform of the interaction potential \cite{schulz93} $V(q)$,
\begin{equation}
\omega (q)= v_{\rm F}|q|\sqrt{1+\frac{V(q)}{\hbar\pi v_{\rm F}}}.
\label{spectrum}
\end{equation}
The particle excitations which change the total electron number are
omitted here. The number of particles is assumed to be constant,
$N_{0}=k_{\rm F}L/\pi $.  The dispersion relation interpolates between
the limit of zero-range interaction ($q\to 0$) where $\omega (k)= v\mid
q \mid$ with the ``charge sound velocity'' $v\equiv v_{\rm F}/g$, and
the limit of non-interacting particles ($q\to \infty$), $\omega (q)=
v_{\rm F} |q|$. The strength of the interaction is defined as
\begin{equation}
\frac{1}{g^{2}}\equiv 1 + \frac{V(q=0)}{\hbar \pi v_{\rm F}}\,.
\label{couplingconstant}
\end{equation}
Non-interacting Fermions correspond to $g=1$, repulsive interaction to
$g< 1$.

The particle density $\rho (x)$ can be written in terms of the phase
variable of the Luttinger model
\begin{equation}
\label{phasevariable}
\vartheta (x) = i\sum_{q\neq 0}\mbox{sgn}(q)\sqrt{\frac{1}{2L\mid q \mid}}
        e^{\varphi (q)-iqx}\left(b^{\dagger}_{q} + b_{-q} \right)
\end{equation}
where
\begin{equation}
{\rm e}^{2\varphi (q)}=\frac{v_{\rm F}\mid q \mid }{\omega (q)}\,.
\end{equation}
With the mean particle density $\rho _{0}=N_{0}/L$ we write
\begin{equation}
\label{density}
\rho (x)\equiv \rho _{0} + \frac{1}{\sqrt{\pi }}\partial_{x}\vartheta (x).
\end{equation}
For later use in the linear response theory, we need the
coupling to the driving voltage $U(x,t)$ and the
current density \cite{sassetti96}. The former is given
by
\begin{equation}
\label{driving}
H_{U}=e\int_{-\infty}^{\infty}dx\rho (x)U(x,t).
\end{equation}
The electric field is $E(x,t)=-\partial_{x}U(x,t)$. The external
voltage is assumed to be given by $\int_{-\infty}^{\infty} dx
E(x,t)\equiv U(t)$. The current operator is defined by using the 1D
continuity equation for the Heisenberg representation of the operators,
\begin{equation}
J(x,t)\equiv -\frac{e}{\sqrt{\pi }}\dot{\vartheta} (x,t)\,.
\label{current}
\end{equation} 

\subsection{The interaction potential}

In order to obtain the dispersion relation explicitly, we need a
specific model for the interaction. Since we want eventually
to draw some conclusions on quantum wires we start from a
3D screened Coulomb interaction
\begin{equation}
\label{thomasfermi}
V(r)=V_{0}\frac{e^{-\alpha r}}{r},
\end{equation}
with $V_{0}=e^{2}/4\pi \varepsilon \varepsilon _{0}$ and project on the
quasi-1D states of the quantum wire.

For the latter, we assume a parabolic confining potential in the
$y$- and $z$-directions independent of $x$. The corresponding states
are ($\zeta =\sqrt{y^{2}+z^{2}}$), 
\begin{equation}
\psi _{k}(x,\zeta )=\frac{e^{ikx}}{\sqrt{L}}\varphi (\zeta )\,.
\label{wirestates}
\end{equation} 
In the following, we assume for the confining wave function
\begin{equation}
\varphi (\zeta )=\sqrt{\frac{2}{\pi d^{2}}}e^{-\frac{\zeta ^{2}}{d^{2}}},
\label{groundstate}
\end{equation}
where $d$ represents the ``diameter'' of the wire.

We obtain the effective interaction potential for the motion in the
$x$-direction from the matrix elements of (\ref{thomasfermi}) in the
state (\ref{groundstate}) by performing the integrations with respect
to $y$ and $z$,
\begin{equation}
\label{effpot}
V(x)=-\frac{2V_{0}}{\alpha d^{2}}\int_{0}^{\infty}
        d\zeta e^{-\frac{\zeta ^{2}}{\alpha ^{2}d^{2}}}
        \frac{d}{d\zeta }
        \left[e^{-\sqrt{\alpha ^{2}x^{2}+ \zeta ^{2}}}\right]\,.
\end{equation}
Its Fourier transform is
\begin{equation}
V(q)=V_{0}e^{\frac{d^{2}}{4}\left(q^{2}+\alpha ^{2}\right)}
        E_{1}\left(\frac{d^{2}}{4}\left(q^{2}+\alpha ^{2}\right)\right).
\label{ft(effpot)}
\end{equation}
The function $E_{1}$ is the exponential integral \cite{abra72}.

Two limiting cases are of particular interest. When
$z\gg 1$, $E_{1}(z)\approx\exp{(-z)}/z$. Thus, for $\alpha d\gg 1$,
\begin{equation}
V(q)=
\frac{4V_{0}}{d^{2}}\frac{1}{q^{2}+\alpha ^{2}}.
\label{ftlutt}
\end{equation}
This is the Fourier transform of
\begin{equation}
V(x)=\frac{1}{2}V_{L}\alpha e^{-\alpha |x|}.
\label{lutt}
\end{equation}
For $\alpha \to \infty$ but with $V_{L}\equiv 4V_{0}/\alpha
^{2}d^{2}=\mbox{const}$, this is $V_{L}\delta (x)$, the zero-range interaction
with the strength $V_{L}$ of the conventional Luttinger liquid.

When $\alpha d\ll 1$ we still get the above result (\ref{ftlutt}) 
as long as $qd\gg 1$  which implies that in $x$-space $V(x\to 0) $
behaves as (\ref{lutt}) but with $V(x=0)=\sqrt{\pi }V_{0}/d$. For $z\to
0$, $E_{1}(z)\approx  -\ln{z}$ so that
for $qd\ll 1$,
\begin{equation}
\label{ftcoul}
V(q)\approx - V_{0}\ln{\left(\left(\alpha^{2} + q^{2}\right)
d^{2}\right)}.
\end{equation}
This is the same behavior as obtained by starting from the
1D-equivalent of the Coulomb interaction \cite{cuni96,schulz93,sarma92}
implying the interaction falls off as $x^{-1}$ in space.

In many of the quantum wire experiments metallic gates are present in
some distance, say $D$, from the wire (diameter $d\ll D$). In order to
discuss the changes in the interaction induced by the presence of the
gates we can consider an infinite metal plate parallel to the wire.
This changes the interaction potential according to
\begin{equation}
V_{D}(r)=V(r)-V(|\vec{r}+\vec{D}|),
\end{equation}
due to the presence of the mirror charge.
It is clear that this influences the results only when $\alpha ^{-1}\ge
D$. Assume then $\alpha = 0$. The cutoff of the Coulomb tail of the
potential is in this limit given by $D$ instead of $\alpha ^{-1}$. The
results to be discussed below for $\alpha d\ll 1$ apply also for
this limit with $\alpha $ replaced by $D^{-1}$. 

\subsection{Results for the dispersion law}

\begin{figure}
%\vspace{6cm}
%\centerline{\epsfig{file=dr_g01ln.eps, angle=270, width=\linewidth}}
\centerline{\epsfig{file=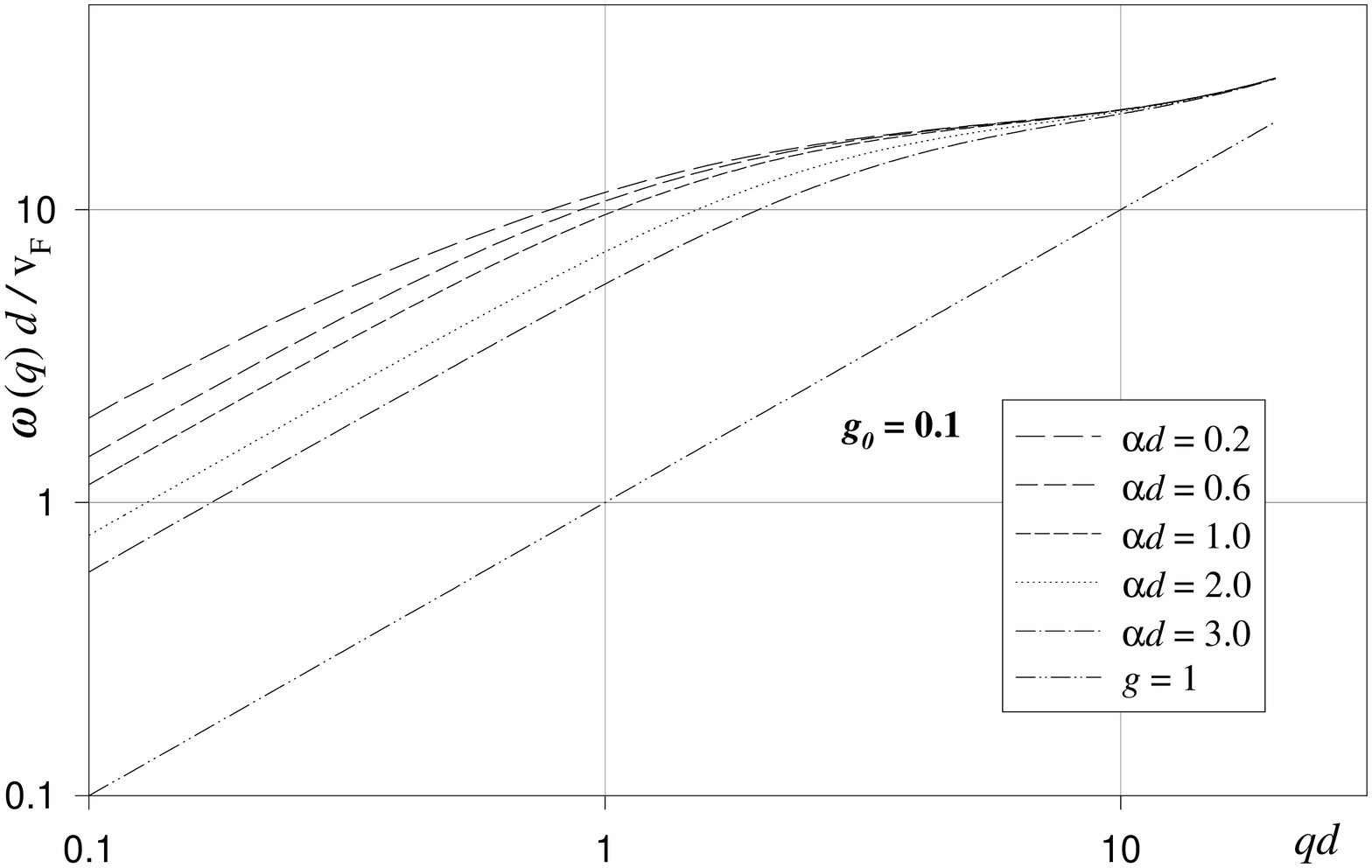, width=\linewidth}}
\caption[1]{Double logarithmic plot of the dispersion relation $\omega (q)$
of the Luttinger model with $g_{0}=0.1$ and different ranges $\alpha d$
($\alpha $ inverse potential range in position space, $d$ diameter of
quantum wire).}
\label{dispersion}
\end{figure}

Results for the dispersion are shown in Fig.~\ref{dispersion} 
for interaction parameter $g_{0}=0.1$ ($V_{0}\equiv \hbar\pi
v_{\rm F}(g_{0}^{-2}-1)$)
and various $\alpha d$. 

There is a crossover between the
interacting and the non-interacting limits for $q\to 0$ and $q\to
\infty$, respectively, at the intermediate wave number $q_{\rm p}$
corresponding to the characteristic frequency $\omega _{\rm p}$. It is
related to the finite range of the interaction in the wave-number
space. For
zero-range interaction the dispersion becomes linear, and
$\omega (q)=v_{\rm F}q/g$. 
\begin{figure}
%\vspace{5cm}
%\centerline{\epsfig{file=qp_z1.eps, angle=270, width=\linewidth}}
\centerline{\epsfig{file=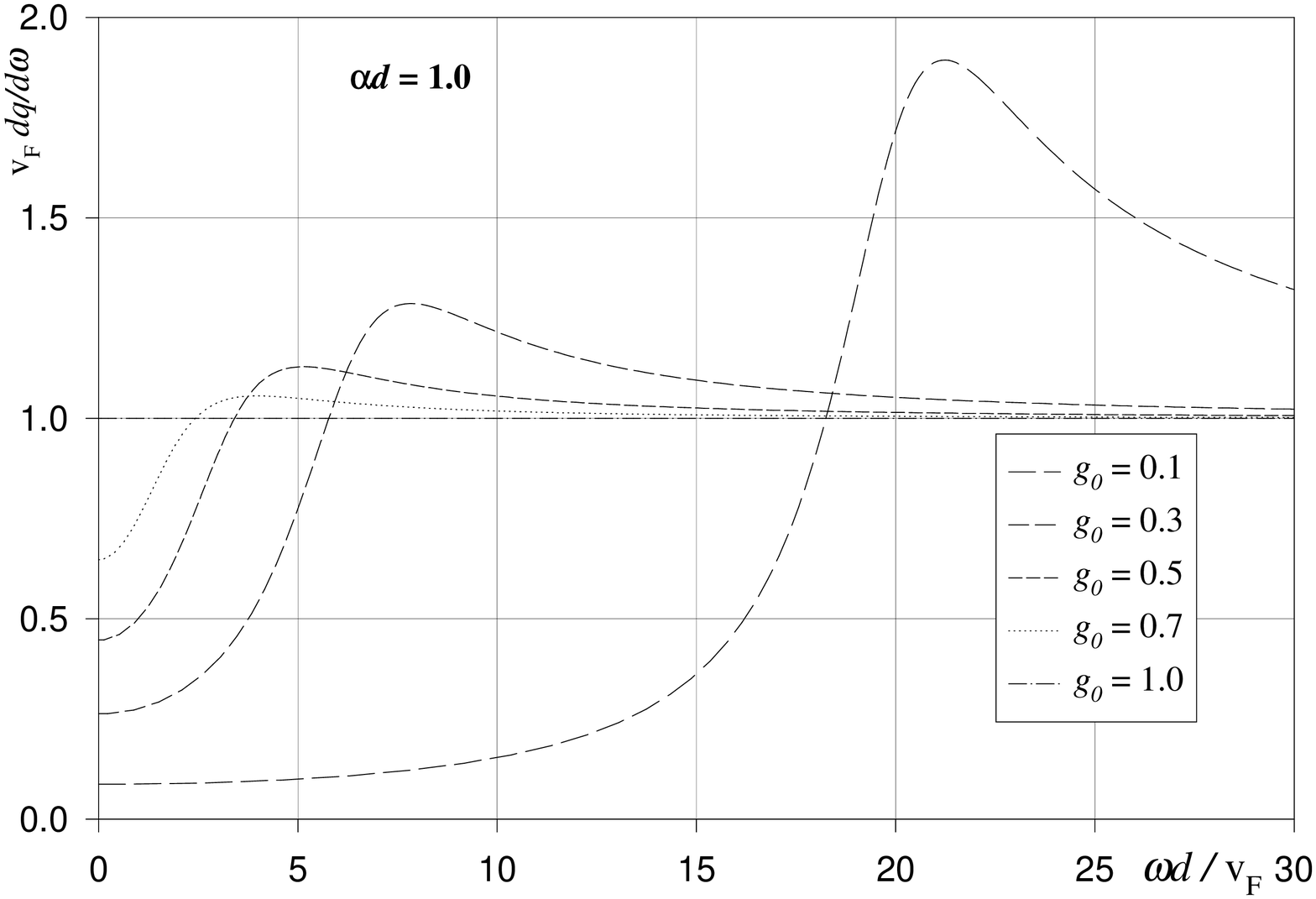, width=\linewidth}}
\caption[2]{The density of pair excitations,
$dq/d\omega $ for $\alpha d=1$, different $g_{0}$.}
\label{plasmonpeak}
\end{figure}  
Fig.~\ref{plasmonpeak} 
shows the excitation
density ${\rm d}q/{\rm d}\omega $ for various $g_{0}$ and $\alpha d=1$.

\begin{figure}
%\vspace{5cm}
%\centerline{\epsfig{file=op_pl.eps, angle=270, width=\linewidth}}
\centerline{\epsfig{file=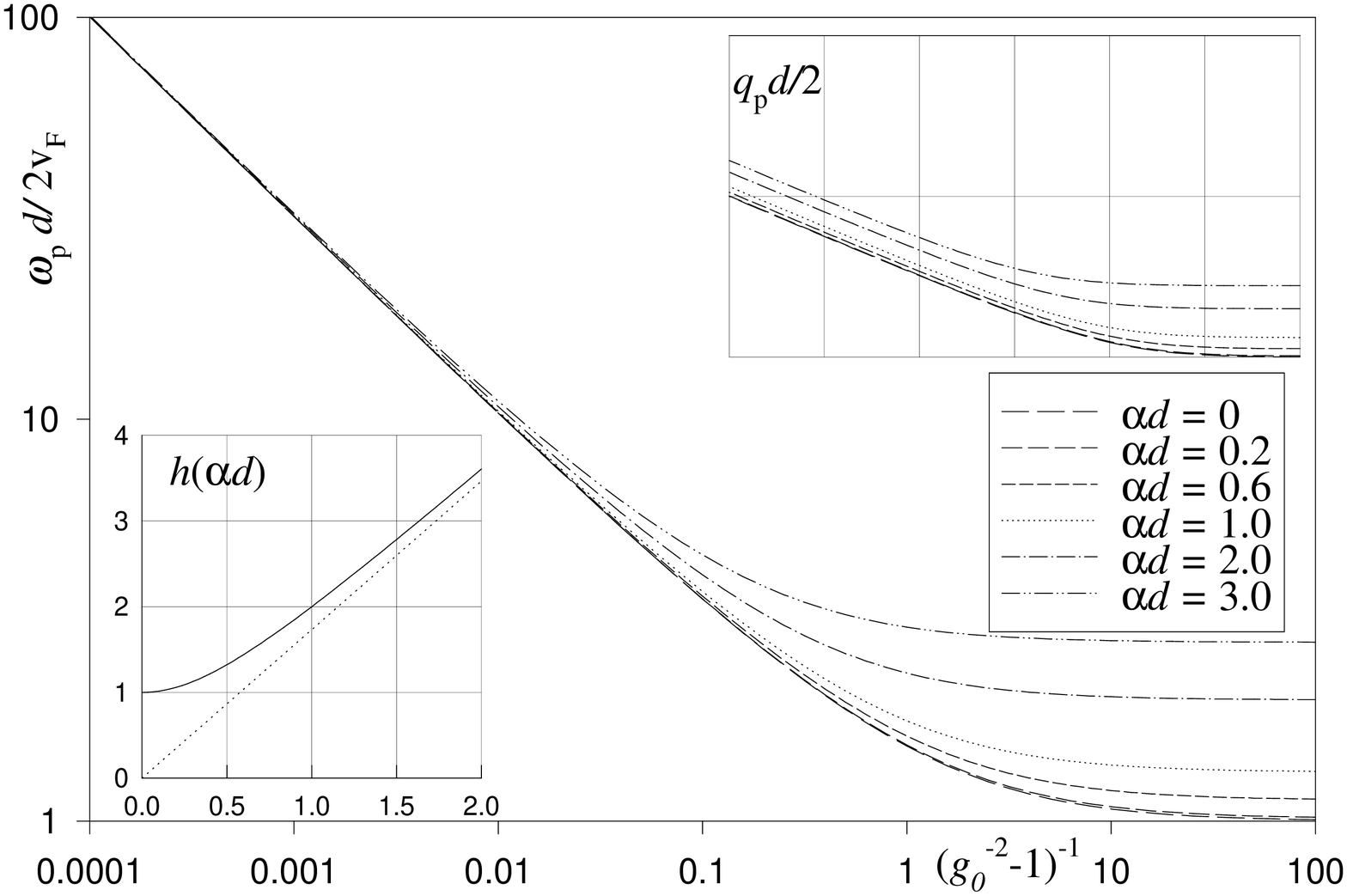, width=\linewidth}}
\caption[3]
{The resonance frequency $\omega _{\rm p}$
and the corresponding wave
number, $q_{\rm p}$ (insert top right) as a function of the strength of the
interaction, $(g_{0}^{-2}-1)^{-1}$ of the
Luttinger liquid with interaction of finite
range as indicated. Insert bottom left: scaling function $h(\alpha d)$.}
\label{plasmon}
\end{figure}

The frequency $\omega _{\rm p}$ and the corresponding wave number
$q_{\rm p}$ as a function of $(g_{0}^{-2}-1)^{-1}$
are shown in Fig.~\ref{plasmon} 
for various $\alpha d$. 

For a broad range
of $g_{0}$ the frequency $\omega _{\rm p}$ and
$q_{\rm p}$ decay as $g_{0}^{-1}$ and
$g_{0}^{-1/2}$, respectively.
The data for $\omega _{\rm p}$ obey the scaling law 
$(\beta_0\equiv g_{0}^{-2}-1)$
\begin{equation}
\omega _{\rm p}(\beta _{0};\alpha d)=
        \omega _{\rm p}(\beta _{0}h^{-2}(\alpha d);0)
                h(\alpha d)\,.
\label{scaling}
\end{equation}
The scaling function $h(\alpha d)$ is shown in the bottom left insert of
Fig.~\ref{plasmon} and it is proportional to the limit of $\omega _{\rm 
p}$ for infinitesimally small interaction.

\section{Linear response}

In this section we outline the calculation of the conductance with
linear response theory.

\subsection{The conductivity}

Using the above current, (\ref{current}), one
obtains the complex conductivity, $\sigma (x-x';t-t')$ which relates the
current at a given point $x$ at time $t$ with the driving electric field
$E_{\rm ext}(x',t')$,
\begin{equation}
J(x,t)=\int_{-\infty}^{\infty}
        {\rm d}x'\int_{-\infty}^{t}
        {\rm d}t'\sigma (x-x';t-t')E_{\rm ext}(x',t')\,.
\label{linearesp}
\end{equation}
While the non-locality of the conductivity is unimportant in the
dc-limit, it is crucial for time-dependent transport. 
By assuming that
the electric field is concentrated only near a given point, say
$x'=x_{0}$, we get
\begin{equation}
J(x,t)=\int_{-\infty}^{t}{\rm d}t'\sigma (x-x_{0};t-t')U_{{\rm ext}}(t')
\end{equation}
with the voltage drop $U_{{\rm ext}}(t')=\int{\rm d}x'E_{\rm ext}(x',t')$
dropping only near $x_{0}$. We see that $\sigma (x-x_{0},t-t')$ plays
then the role of a ``conductance'' that relates the current at some
point $x$ in the system with a voltage dropping at some other point. By
using near-field microscopy, one could possibly perform such a
non-local experiment. However, it seems to us that it is very hard to
measure the current locally in a quantum wire, especially in the region
of high frequency. 

By Fourier transformation, (\ref{linearesp}) is equivalent to
\begin{equation}
J(q,\omega )=\sigma (q,\omega )E_{\rm ext}(q,\omega )\,.
\label{fouriertrans}
\end{equation}
The conductivity kernel can be expressed either by the
current-current correlation function or, by using the
continuity equation, as
\begin{equation}
\sigma (q,\omega )\equiv \sigma _{1}(q,\omega)+
                {\rm i}\sigma _{2}(q,\omega)
=\frac{-{\rm i}\omega e^{2}}{q^{2}}R(q,\omega )\,,
\label{chargecharge}
\end{equation}
with the charge-charge correlation function
\begin{equation}
R(x,t)=-\frac{{\rm i}}{\hbar}\Theta (t)\langle [\rho (x,t),\rho (0,0)]\rangle\,.
\end{equation}
Here, $\langle\cdots\rangle \equiv
\mbox{Tr}\left(\exp{(-\beta H)}\cdots\right)/
\mbox{Tr}\exp{(-\beta H)}$
is the usual thermal average (temperature $\propto \beta ^{-1}$).
By using the expression (\ref{density}) we find the exact result
\begin{equation}
R(q,\omega )=\frac{v_{\rm F}}{\hbar\pi }
                \frac{q^{2}}{\omega ^{2}(q)-(\omega
                +{\rm i}0^{+})^{2}}\,.
\label{r}
\end{equation}

\subsection{Analogy with a Brownian particle}

A most remarkable feature of the above result becomes transparent by
applying the imaginary-time path integral approach. With this, the
time-dependent non-linear response to an electric field of arbitrary
spatial shape of a Luttinger system was calculated recently
\cite{sassetti96}. It was found that the average of the phase field
(\ref{phasevariable}) obeys the equation of motion of a Brownian
particle with mass $M\to 0$
\begin{equation}
\label{brownian}
M\ddot{\vartheta}(x,t) + \int_{-\infty}^{t}{\rm d}t'
        \gamma (t-t')\dot{\vartheta}(x,t')
= {\cal L}(x,t)
\end{equation}
subject to an effective external force with the Fourier
transform
\begin{equation}
{\cal L}(x,\omega )=-\frac{e}{\sqrt{\pi }}\frac{1}{\sigma (x=0,\omega )}
        \int_{-\infty}^{\infty}{\rm d}x' E(x',\omega )\sigma (x-x',\omega ),
\end{equation}
and a damping term $\gamma (t)$ which is given by the non local conductivity.
The  Fourier transform of $\gamma (t)$ is
\begin{equation}
\gamma (\omega )=\frac{e^{2}}{\pi }\frac{1}{\sigma (x=0,\omega )}.
\label{damping}
\end{equation}
>From the solution of the equation of motion the current is found,
\begin{equation}
J(x,t)=\int_{-\infty}^{\infty}{\rm d}x'\int_{-\infty}^{t}{\rm d}t'
        \sigma (x-x',t-t')E(x',t')\,.
        \label{browniansigma}
\end{equation}
Linear response is exact for the Luttinger liquid. 

\subsection{The driving electric field}

In this section, we investigate the influence of screening on the response to
an external electric field. In particular, we discuss the dc-limit and show
that the two limits $\omega \to 0$ and $q\to 0$ cannot be interchanged. It
will turn out that for the Luttinger model, where linear response is exact,
one can use the external field for the calculation of the current. The results
will be used to derive absorptive and reactive conductances.

The dielectric response function, which describes the dynamical
screening of a charge, is defined as
\begin{equation}
\varepsilon ({q,\omega })=
\frac{U_{\rm ext} (q,\omega )}{U_{\rm tot}(q,\omega )}\,,
\label{dk}
\end{equation}
where
\begin{equation}
U_{\rm tot}(q,\omega )=U_{\rm ext}(q,\omega ) + U_{\rm sc}(q,\omega )
\label{potential}
\end{equation}
is the total potential and $U_{\rm ext}$, $U_{\rm sc}$
are the external and screening potential, respectively.
Within the linear screening model, the
dielectric response function can be written in terms of the
charge-charge correlation function as
\begin{equation}
\frac{1}{\varepsilon (q,\omega )}= 1 - V(q)R(q,\omega )\,.
\end{equation}
By using the result (\ref{r}) for $R(q,\omega )$ we find
the explicit expression
\begin{equation}
\varepsilon (q,\omega )=\frac{\omega ^{2}(q)-\omega ^{2}}
        {\omega^{2}_{0}(q)-\omega ^{2}}\,,
\label{dklutt}
\end{equation}
with the dispersion of the non-interacting electrons $\omega _{0}(q)$.

Equations (\ref{dk}) and (\ref{potential}) imply that
\begin{eqnarray}
E_{\rm tot}(q,\omega )&=& E_{\rm ext}(q,\omega )
        \left[1-V(q)R(q,\omega )\right]\nonumber\\
        &\equiv&E_{\rm ext}(q,\omega )F(q,\omega )\,.
\end{eqnarray}
Using the conductivity (\ref{chargecharge}) with
(\ref{r}) and comparing with (\ref{dklutt}) we see that
\begin{equation}
\sigma (q,\omega )=\frac{\sigma _{0}(q,\omega )}{\varepsilon (q,\omega )}.
\end{equation}
Here, $\sigma _{0}$ is the conductivity of the non-interacting
electrons. This implies that
\begin{equation}
E_{\rm tot}(q,\omega )\sigma _{0}(q,\omega )=
E_{\rm ext}(q,\omega )\sigma (q,\omega )\,.
\end{equation}
If we calculated the conductivity from the response to the total
field, the conductivity would turn out to be that of non-interacting
electrons.

This result is true as long as one can use linear
screening. It implies also that the voltage drop at frequency $\omega $ is
the same for both fields
\begin{equation}
U_{\rm tot}(\omega )=\int_{-\infty}^{\infty}{\rm d}xE_{\rm tot}(x,\omega )=
        U_{\rm ext}(\omega )
\end{equation}
since for any finite non-zero frequency $\varepsilon (q\to 0, \omega )=1$. 

For a monochromatic external field, 
\begin{equation}
E_{\rm ext}(q,t)=E_{\rm ext}(q)\cos \omega t\,,
\end{equation}
one finds the result
\begin{eqnarray}
E_{\rm tot}(q,t)&=&E_{\rm ext}(q)
        \left[{\cal R}{\rm e}F(q,\omega )\cos \omega t\right.\nonumber\\
&&\left.        \qquad + {\cal I}{\rm m}F(q,\omega )\sin \omega t \right]
\label{fields}
\end{eqnarray}
which means that there is a phase shift between the total and the
external field.

A final remark concerns the static limit. While we have for any non-zero
frequency $\varepsilon (q\to 0, \omega )=1$, we find for $\omega =0$
\begin{equation}
\lim _{q\to 0}\varepsilon (q,0)=
        \lim _{q\to 0}\frac{\omega^{2}(q)}{\omega _{0}^{2}(q)}=\frac{1}{g}\,.
\end{equation}
By inserting the dispersion relation of the Luttinger model into
(\ref{r}) one obtains
\begin{equation}
E_{\rm tot}(q)=E_{\rm ext}(q)\frac{1}{1+V(q)/\hbar v_{\rm F}\pi }\,.
\label{renorm}
\end{equation}

The limits $\omega \to 0$ and $q\to 0$ cannot be inter-changed. The latter
result has been used recently, in order to explain that in quantum wires the
dc-conductance is not renormalized by the interaction
\cite{kawabata,finkelstein} (see also (\ref{renorm})).  We are discussing here
frequency dependent properties. Thus, we always obtain for small frequencies a
conductance that is renormalized by the interaction since we consider the
limit of infinite system length.

\subsection{Absorptive and reactive conductances}

Since it is very difficult to detect experimentally the non-local
conductivity some average has to be performed. One possibility is to
use the absorbed power, $P(t)$, in order to define the conductance.
This appears to be a natural choice if one wants to describe 
infrared or microwave experiments,
\begin{eqnarray}
P(t)&=&\int_{-\infty}^{\infty}{\rm d}x J(x,t)E_{\rm tot}(x,t)\nonumber\\
        &=&\frac{1}{2\pi }\int_{-\infty}^{\infty}
        {\rm d}q J(q,t)E_{\rm tot}(-q,t)\,.
\label{power}
\end{eqnarray}
We define the average
\begin{equation}
\overline{P}=\lim_{T\to \infty}\frac{1}{T}\int_{0}^{T}{\rm d}tP(t)\,.
\label{average}
\end{equation}
Using the Laplace transform
\begin{equation}
P(s)=\int_{0}^{\infty}{\rm d}t e^{-st}P(t)
\end{equation}
we obtain
$\overline{P}=\lim_{s\to 0}sP(s)$.
The {\em absorptive conductance} can then be defined by
\begin{equation}
\Gamma _{1}=\frac{\overline{P}}{\overline{U^{2}}_{\rm ext}}
\label{gamma1}
\end{equation}
with $U_{\rm ext}(t)=\int {\rm d}x E_{\rm ext}(x,t)$.
It is independent of the amplitude of the external field but depends
on its shape in space and time. Physically, it is the absorption
constant for electromagnetic radiation.
Using (\ref{fields}) one can show that the absorbed power for a
monochromatic time dependence is the same for both fields due to the
time average. We obtain
\begin{equation}
\overline{P}_{\omega}=
        \frac{1}{4\pi }\int_{-\infty}^{\infty}
        {\rm d}q\,{\cal R}{\rm e}\sigma (q,\omega )
        \mid E_{\rm ext}(q)\mid^{2}\,.
\end{equation}

With (\ref{r}) we find for the real
part of the conductivity
\begin{equation}
{\cal R}{\rm e}\sigma (q,\omega )=\frac{v_{\rm F}e^{2}}{2\hbar}
\left[\delta (\omega - \omega (q)) + \delta (\omega  + \omega (q))\right]\,.
\end{equation}
This gives the expression 
\begin{equation}
\Gamma_1(\omega)=v_{\rm F}\frac{e^2}{h}L(q(\omega))
\frac{\mbox{d}q}{\mbox{d}\omega}
\label{realpart}
\end{equation}
with the Fourier transformed
auto-correlation function of the external electric field
\begin{equation}
\label{autocorrelation}
L(q) \equiv \frac{1}{2 \overline{U^{2}}_{\rm ext}}
        \left|\int_{-\infty}^{\infty}{\rm d}x
        e^{iqx}E_{{\rm ext}}(x)\right|^{2}\,.
\end{equation}

By a Kramers-Kronig transformation, we can define also a {\rm reactive
conductance} 
\begin{eqnarray}
\Gamma _{2}(\omega )&=&\frac{1}{\pi }
        {\cal P}\int _{-\infty}^{\infty}
        {\rm d}\omega '\frac{\Gamma _{1}(\omega ')}
        {\omega -\omega '}\nonumber\\
        &=&\frac{1}{4\pi \overline{U^{2}}_{\rm ext}}
        \int_{-\infty}^{\infty}{\rm d}q\,{\cal I}{\rm m}\sigma (q,\omega )
        \mid E_{\rm ext}(q)\mid^{2}\,,
\end{eqnarray}
it contains information about phase shifts.

For a zero-range interaction the conductance becomes
\begin{equation}
\Gamma _{1}(\omega ) = \frac{ge^{2}}{h}\,
        L\left(\frac{g\omega }{v_{\rm F}}\right),
        \label{freeconductance}
\end{equation}
the same as without interaction \cite{velicky89}, except
for the renormalization of the prefactor and the
Fermi velocity with the interaction
strength $g$ and $g^{-1}$, respectively.
In the general case of an interaction potential of finite range we get
asymptotically ($\omega \to \infty$)
\begin{equation}
\Gamma _{1}(\omega ) \approx
\frac{e^{2}}{h}L\left(\frac{\omega }{v_{\rm F}}\right).
\end{equation}
This reflects that for large $q$ the dispersion
is not influenced by an interaction of a finite range.

\begin{figure}
%\vspace{5cm}
%\centerline{\epsfig{file=gamma12n.eps, angle=270, width=\linewidth}}
\centerline{\epsfig{file=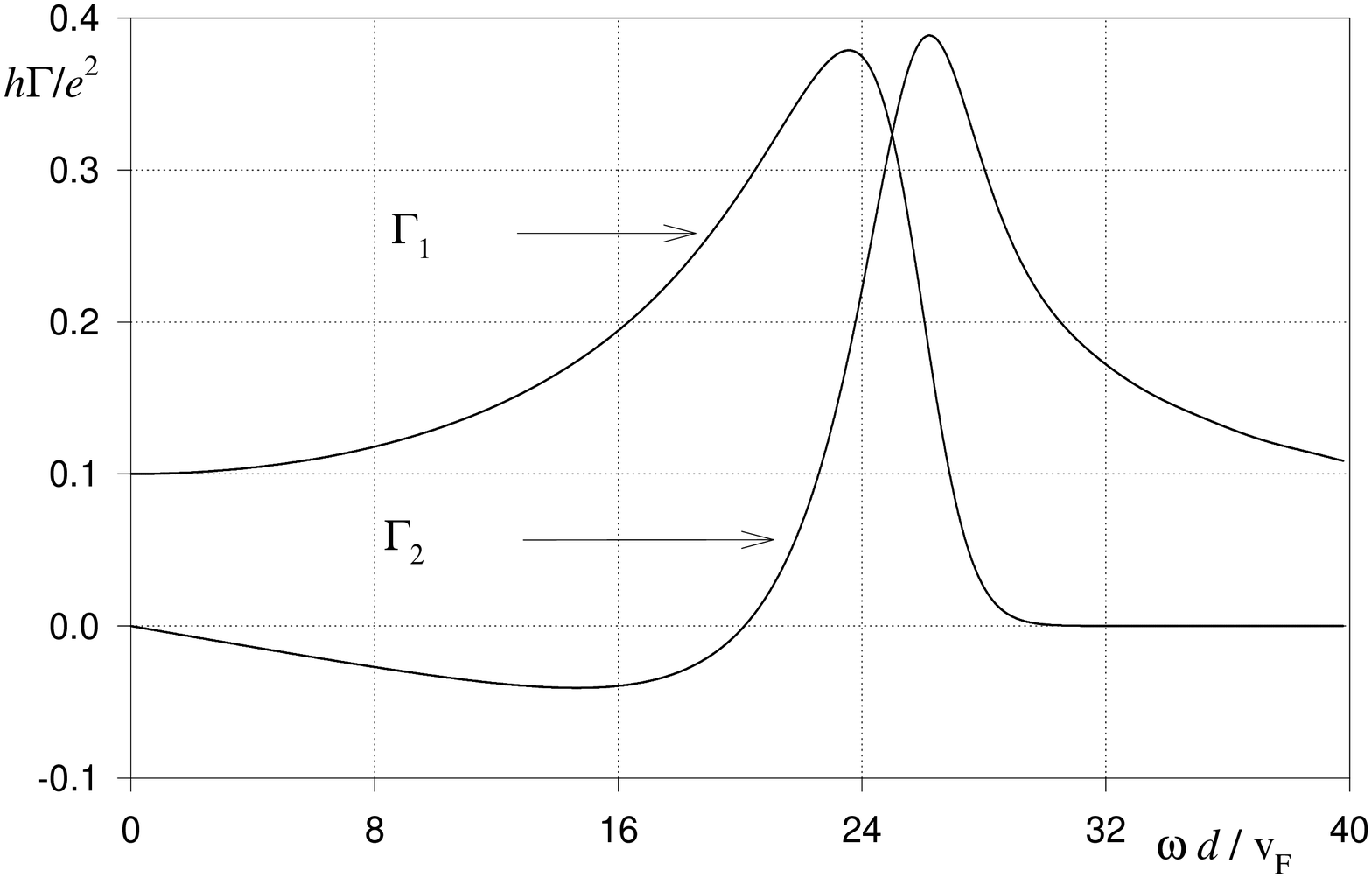, width=\linewidth}}
\caption[4]{Real and imaginary part,
$\Gamma_{1} (\omega )$ and $\Gamma _{2} (\omega )$,
of the
ac-conductance of a Luttinger wire with finite range interaction;
$\alpha d=1$, $g=0.1$,
range of the electric field $\ell /d = 1/4$.}
\label{accond}
\end{figure}

A most important feature of the result (\ref{realpart}) is the
factorization into a part that depends only on the internal properties
of the interacting electron system, ${\rm d}q/{\rm d}\omega $, and a
part that contains the information about the shape of the electric
field. Only in the limit of vanishing frequency, the shape of the
latter is unimportant \cite{velicky89,ponomarenko}. In general, the
ac-response depends on the spatial properties of the electric field
\cite{sassetti96} which is certainly determined by the interactions. 

Most remarkably, the temperature does not enter the result, although
$T\neq 0$ was assumed in the derivation. This is due to the
linearization of the spectrum. As long as this assumption is justified,
the response of the Luttinger liquid is independent of temperature.

A typical result for $\Gamma (\omega )$ is shown in Fig.~\ref{accond}.
The Fourier transform of the electric field has been assumed to be a
Gaussian $E(x)=E_{0}\exp{(-2x^{2}/\ell^{2})}$. If the range of the
electric field is zero, the zero of $\Gamma _{2}(\omega )$, $\omega
^{*}_{\rm m}$, and the position of the maximum of $\Gamma_{1}(\omega
)$, $\omega _{\rm m}$, do not agree. However, as soon as $\ell$ is
finite, $\omega _{\rm m}\approx \omega^{*}_{\rm m}$ for a wide region of
parameters.

In the Coulomb case, $\alpha =0$, $\Gamma _{1}\propto (\ln{\omega
})^{-1/2}$ for $\omega \to 0$, due to the logarithmic singularity of
the dispersion for $q\to 0$.

\section{``Quantum impedances''}

In this section, we analyze the results for the complex conductance
presented above with respect to ``resistive'', ``capacitive'' and
``inductive'' behavior. We compare the Luttinger system with an 
equivalent classical circuit.

\subsection{Impedance network}

\begin{figure}
%\vspace{5cm}
%\centerline{\epsfig{file=circ.eps, width=\linewidth}}
\centerline{\epsfig{file=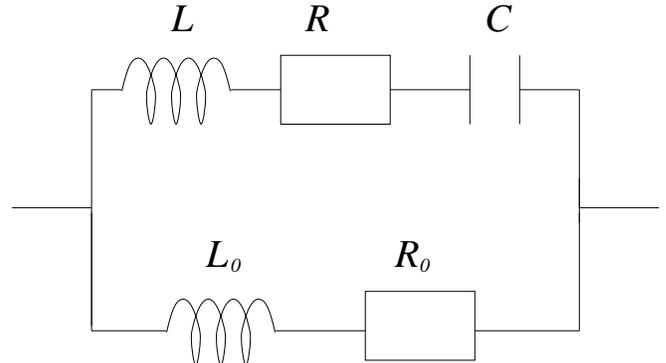, angle=270, width=\linewidth}}
\caption[5]{Classical circuit for simulating the frequency behavior of the
complex conductance.}
\label{circuit}
\end{figure}

Our system of interacting electrons shows a resonance in the
ac-conductance. It can be useful to consider a circuit of
capacitances, inductances and resistances, in order to simulate the
frequency behavior. The circuit shown in  Fig.~\ref{circuit} contains
the minimum set of elements that are necessary for reproducing both the
low-frequency behavior and near the resonance. 
Its complex impedance
$Z(\omega )$ is given by
\begin{equation} \label{impedance}
Z^{-1}(\omega )=\frac{{\rm i}\omega C}{1+{\rm i}\omega RC - \omega
        ^{2}LC} 
+\frac{1}{R_{0}+{\rm i}\omega L_{0}}\,.
\end{equation}
The resistance $R_{0}$ is fixed to be the 
resistance at zero frequency, $h/ge^{2}$. The circuit shows a resonance near 
$\omega_{0}=(LC)^{-1/2}$ with a width depending on $R$.

At low frequency, the real and imaginary parts of $Z^{-1}(-\omega )$, $\Gamma
_{1}(\omega)$ and $\Gamma _{2}(\omega)$, respectively, behave as
\begin{equation}
\Gamma_{1}(\omega )=R_{0}^{-1} +\gamma_{1} \omega ^{2},\quad
\Gamma_{2}(\omega )= \gamma_{2}\omega
\label{smallomega}
\end{equation}
with
\begin{equation}
\gamma _{1}=RC^{2}-\frac{L_{0}^{2}}{R_{0}^{3}},\quad
\gamma _{2}=-C+ \frac{L_{0}}{R_{0}^{2}}\,.
\label{gammas}
\end{equation}
The circuit is defined to behave ``capacitively'' if simultaneously
$\gamma _{1}>0$ and $\gamma _{2}<0$. If simultaneously $\gamma _{1}<0$
and $\gamma _{2}>0$ the circuit is clearly ``inductive''. If $\gamma
_{2}=0$, i.e. there is no phase shift between current and voltage,
$\gamma _{1}$ indicates capacitive or inductive behavior depending on
whether $R/R_{0}$ is larger or smaller than 1, respectively. Note that
until ${\rm O}(\omega ^{2})$ the inductance
$L$ does not play any role. Whether the
circuit behaves capacitively or inductively near the resonance is
therefore to a certain extent independent of its behavior at small
frequency. 

\subsection{Low frequency}

\begin{figure}
%\vspace{5cm}
%\centerline{\epsfig{file=trfa.eps, angle=270, width=\linewidth}}
\centerline{\epsfig{file=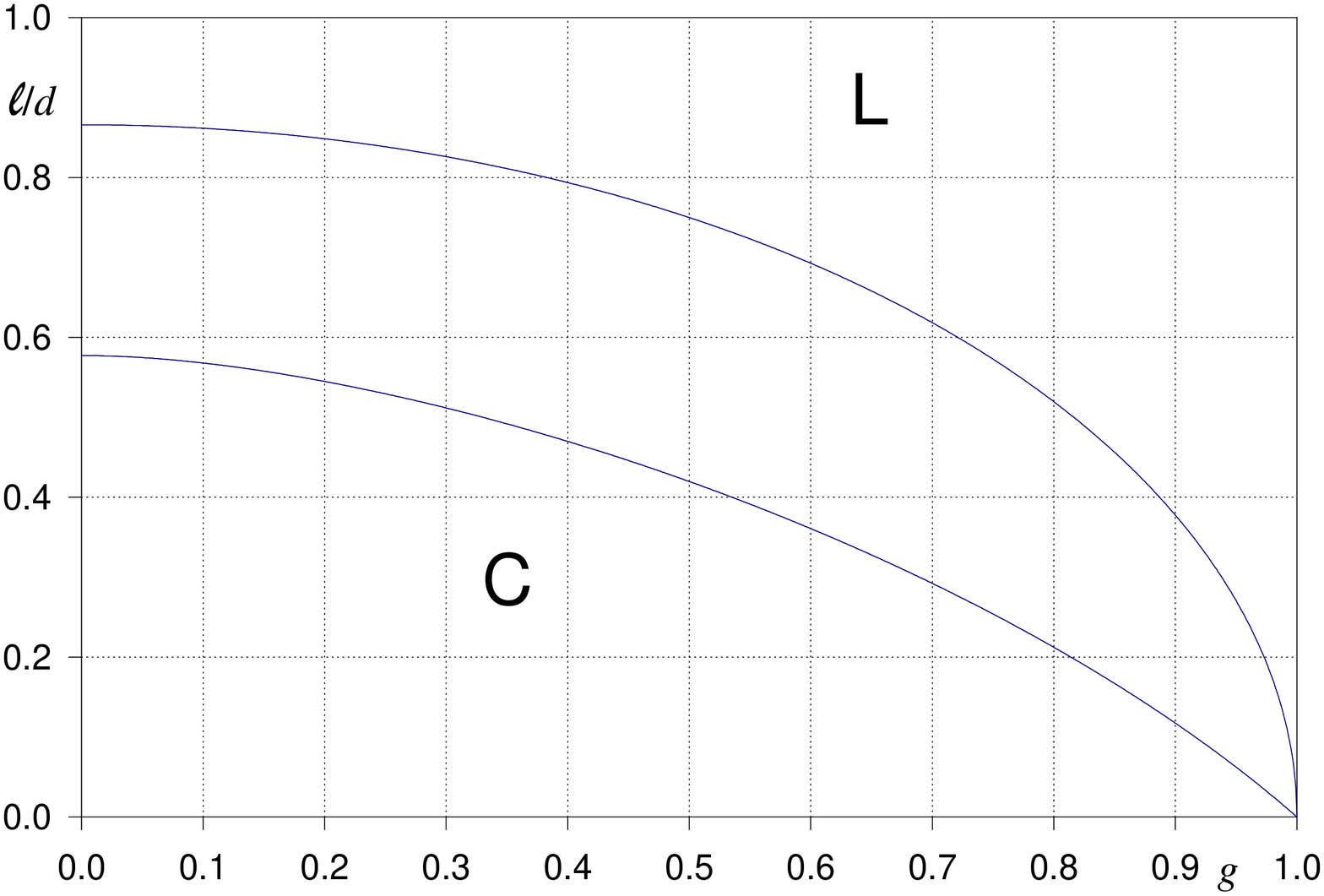, width=\linewidth}}
\caption[7]{``Phase trajectories'' in the $(\ell/d)$-$g$ plane
separating capacitive from inductive
behaviors of the Luttinger liquid for $\alpha d=1$.}
\label{trajectory}
\end{figure}

We have extracted the parameters $\gamma _{1}$ and $\gamma _{2}$ (cf.
(\ref{gamma1low}) and (\ref{gamma2low})) which characterize the low-frequency
behavior of the conductance from the ac-conductance of the Luttinger liquid.
\begin{figure}
%\vspace{5cm}
%\centerline{\epsfig{file=sigma12.eps, angle=270, width=\linewidth}}
\centerline{\epsfig{file=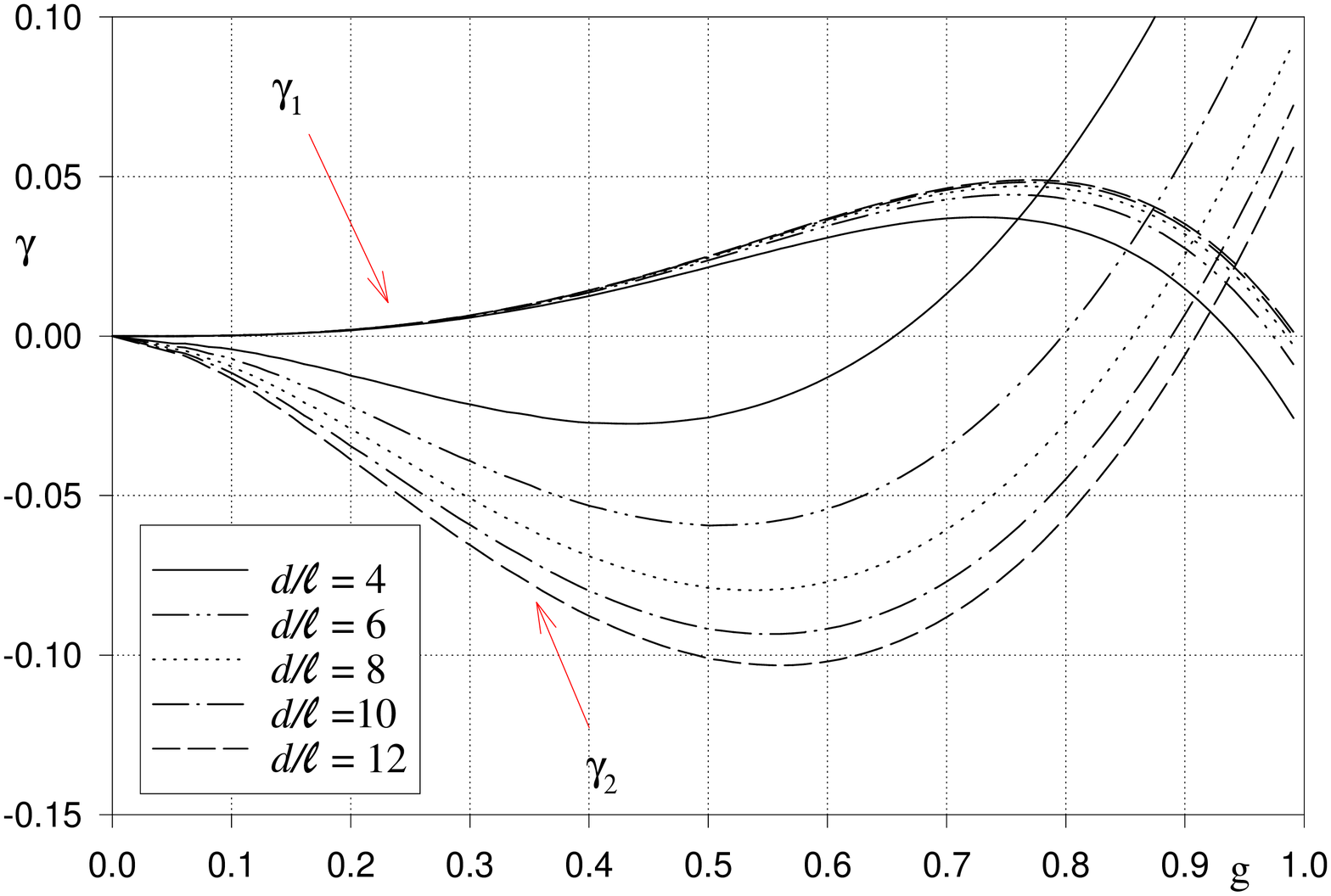, width=\linewidth}}
\caption[6]{The parameters $\gamma _{1}$ and $\gamma _{2}$ (in units of
$e^{2}d^{2}/hv_{\rm F}^{2}$ and $e^{2}d/hv_{\rm F}$, respectively) which
characterize the low frequency behavior of the ac-conductance of the
Luttinger liquid as a function of the interaction parameter $g$ for
different ranges of the electric field, $\ell$.}
\label{figgammas}
\end{figure}
By assuming, as above, the electric field to be the Gaussian distributed, we
obtain the explicit results (in units of $e^{2}/h$ with $V(0)=V(q=0)$)
\begin{eqnarray}
\gamma _{1}&=&\frac{g^{3}d^{2}}{4v_{\rm F}^{2}}
\left[\frac{3(1-g^{2})}{2}\left(\frac{4V_{0}}{\alpha ^{2}d^{2}V(0)}-
1\right)
-\frac{\ell^{2}}{d^{2}}\right]\nonumber\\
&&\label{eqgamma}\\
\gamma _{2}&=&-\frac{2v_{\rm F}}{\pi }\int_{0}^{\infty}{\rm d}q
e^{-q^{2}\ell^{2}/4}\left[\frac{1}{\omega ^{2}(q)}-
\frac{1}{\omega _{g}^{2}(q)}\right]
+\frac{g^{2}\ell}{v_{\rm F}\sqrt{\pi }}\nonumber
\end{eqnarray}
where $\omega _{g}(q)=v_{\rm F}|q|/g$.
Figure~\ref{figgammas} 
show $\gamma _{1}$ and $\gamma _{2}$ as a function of
$g$. Depending on the range of the field, $\ell$, the behavior changes from
capacitive (small $\ell$, $\gamma _{1}>0$, $\gamma _{2}<0$) to inductive
(large $\ell$, $\gamma _{1}<0$, $\gamma _{2}>0$). 
As functions of $g$, $\gamma
_{1}$ and $\gamma _{2}$ change also signs. Always, this change of sign happens
at a smaller value of $g$ for $\gamma _{2}$. The ``phase separation'' lines
defined by $\gamma _{1}(\ell ,g_{1})=\gamma _{2}(\ell, g_{2})=0$ are shown in
Fig.~\ref{trajectory} for $\alpha d=1$. 
By decreasing $\alpha d$, i.e.
increasing the range of the interaction, the end points of the two
trajectories at $g=0$ are shifted to higher values of $\ell/d$. The region of
capacitive-like behavior increases at the expense of the inductive region for
increasing interaction strength.

>From the behaviors deep in the capacitive and inductive regions, one can
deduct formulas for equivalent inductances and capacitances.  However, these
are not always unique, although their scaling properties with the parameters
of the system are. For instance, in the region denoted by ``L'' in
Fig.~\ref{trajectory} $\gamma _{2}$ is given by the second term in
(\ref{eqgamma}) only. This leads, by comparing with (\ref{gammas}), to define
$L_{0}=h\ell/\sqrt{\pi }e^{2}v_{\rm F}$, independently of the interaction
parameter. Qualitatively the same result is obtained when using the expression
for $\gamma _{1}$. Apart from a different numerical prefactor, the scaling of
the inductance with $\ell$ is the same. This is related to the fact that the
behavior of $\Gamma _{2}$ at low frequency is determined via the
Kramers-Kronig transformation to the behavior of $\Gamma _{1}$ also at high
frequency.  The latter depends strongly on the shape of the field. Only if
this is assumed in order to reproduce the $\omega ^{-2}$-behavior of the
classical circuit, one can expect $\gamma _{1}$ and $\gamma _{2}$ to yield the
same $L_{0}$.

In the capacitive region, where $g\ll 1$, the second
terms on the right hand sides of (\ref{eqgamma}) can be neglected. The
first terms can be used to define an equivalent capacitance $C$ and a
dissipative resistance $R$ by comparing with (\ref{gammas}). For
$\alpha d\ll 1$, in the limit of Coulomb interaction, we obtain from
$\gamma _{2}$
\begin{equation}
C\approx \frac{e^{2}}{h}\frac{4\beta _{0}}{\alpha v_{\rm F}}
\frac{1}{(1-2\beta _{0}\ln{\alpha d})^{2}}\,, 
\label{c0}
\end{equation}
which is independent of the electric field and diverges for infinite
interaction range, $\alpha \to 0$.
By comparing with the first term of
$\gamma_{1}$ we obtain the dissipative resistance
\begin{equation}
R\approx \frac{h}{e^{2}}\frac{3}{32}\frac{1}{\beta _{0}}
\left(1-2\beta_0\ln{\alpha d}\right)^{3/2}\,.
\end{equation} 
For $\alpha d>1$, we find
\begin{eqnarray}
C&\approx &
\frac{e^{2}}{h}\frac{d}{v_{\rm F}}\frac{\beta_{0} }{(z^{2}+
\beta_{0} ^{2})^{3/2}}
\exp{\left[\frac{\beta_{0} +z^{2}}{8}\frac{\ell^{2}}{d^{2}}\right]}
\times\nonumber\\
&&\qquad\qquad\qquad\times \left[1-\Phi\left(\frac{\ell}{2d}
\sqrt{\frac{\beta_{0} +z^{2}}{2}}\right)\right]
\label{c1}
\end{eqnarray} 
with $z=\alpha d/2$
and $\Phi $ the
error function. Here, the capacitance depends on the electric field, for
instance, 
\begin{equation}
C=\frac{e^{2}}{h}C_{1}\frac{g(1-g^{2})}{v_{\rm F}\alpha }
\label{clow}
\end{equation}
with $C_{1}=2$ for $g/\ell \alpha \gg 1$ and $C_{1}=(8\sqrt{2/\pi })(g/
\alpha \ell)$ for $g/\ell \alpha \ll 1$.
The corresponding dissipative resistances are
\begin{equation}
R=\frac{h}{e^{2}}R_{1}\frac{g}{1-g^{2}}
\label{rlow}
\end{equation}
with $R_{1}=3/8$ and $R_{1}=(3\pi /256)(\alpha \ell /g)^{2}$
for $g/\ell \alpha \gg 1$ and $g/\ell \alpha \ll 1$, respectively.

\subsection{Near resonance}

As seen in fig.~\ref{accond} the absorptive conductance shows a resonance near
the frequency $\omega_{\rm p}$. For this, the interaction has to be very
strong, i.e. $g\ll 1$.  Furthermore, the range of the electric field should
not be too large (see below). Due to the scaling (cf. (\ref{scaling})) the
results in the Coulomb and Luttinger limits are closely related. Indeed, in
contrast to the limit $\omega \to 0$, in the Coulomb limit $\alpha d\ll 1$ the
conductance does not show any singular behavior near the resonance frequency
$\omega _{\rm m}$.  Since near the resonance the zero-frequency
resistance does not
play any role, $R_{0}$ and $L_{0}$ are neglected when fitting $C$, $L$ and $R$
to the ac-conductance. The parameters of the circuit can be obtained by
fitting to the resonance frequency, and the width and the height of the
resonance.

For small range of the electric field, $\ell \ll
q_{\rm p}^{-1}$, we find 
\begin{eqnarray}
C&\propto& \frac{e^{2}}{hv_{F}}\beta_{1} ^{3/2}\lambda\,, \\
L&\propto& \frac{h}{v_{\rm F}e^{2}}\beta_{1} ^{1/2}\lambda\,, \\
R&\propto& \frac{h}{e^{2}}\beta_{1} ^{1/2}
\label{intrinsic}
\end{eqnarray}
with $\beta_{1} = g$, $\lambda = \alpha ^{-1}$ and $\beta_{1} = g_{0}$,
$\lambda = d$ for $\alpha d\gg 1$ and $\alpha d\ll 1$,
respectively.

If, on the other hand, the range of the electric field is large, $\ell >
q_{\rm p}^{-1}$, the resonance becomes smaller and very broad. The parameters
of the circuit depend here again on the field range. We find
\begin{eqnarray}
C&\propto&\frac{e^{2}}{hv_{\rm F}}\frac{\lambda^{2}}{\ell}\beta_{1} ^{2}\\
L&\propto&\frac{h}{v_{\rm F}e^{2}}\ell\\
R&\propto&\frac{h}{e^{2}}\frac{1}{\beta_{1} }\frac{\ell^{3}}{\lambda ^{3}}\,.
\label{broad}
\end{eqnarray}
Remarkably, the capacitance obtained here scales in the same way  with
the system parameters as the one obtained in the limit of low
frequency, cf. (\ref{clow}). Also the inductance scales as for $\omega
\to 0$, although $L\neq L_{0}$. The dissipative resistance, however,
scales differently and depends much stronger on $\ell$ than $L$ and
$C$. This reflects, that the dissipative resistance is much more
sensitive to the width and the height of the resonance than $L$ and
$C$. The product of the latter is fixed by the resonance frequency
which is quite insensitive to the range of the electric field in a
broad region of field ranges. If we assume $L=L_{0}$ the scaling
behavior of $C$ near the resonance follows directly from $\omega
_{\rm m}=(LC)^{-1/2}$. Thus, the somewhat astonishing result of this
is that one can identify a region of parameters for the 1D electron
system, in which the scaling properties of $L$ and $C$ are independent
of the frequency, though the numerical prefactors can be different. 

\section{Discussion}

We considered the ac-transport properties of quantum wire with finite
range interaction. The linear response theory was found to be exact,
consistent with earlier work, as a result of the linearization of the
dispersion relation. The dependence of the current on the electric field is
given by the microscopic non-local conductivity. However, the latter is not
very useful when aiming at a description of experiments. What is needed is a
description in terms of externally accessible macroscopic quantities, as for
instance given by 
Ohm's law. Such a relation can also be found in
the present, non-local quantum region. By assuming the electric field to be
non-zero near a given point $x_{j}$ and the current to be detected by a
probe at a point $x_{i}$ one finds $\Gamma (\omega ) \equiv
\Gamma _{ij}(\omega )=\sigma (x_{i}-x_{j},\omega )$.  By generalizing to
several probe positions $x_{1}\ldots x_{p}$ one obtains
\begin{equation}
J_{i}(\omega )=\sum_{j} \Gamma _{ij}(\omega ) U_{j}(\omega )\,.
\end{equation}
Such an approach has been used recently \cite{bc96} in order to generalize the
Landauer dc-approach to finite frequency. In the present work, the non-local
conductances $\Gamma _{ij}(\omega )$ are natural results of the response
theory when suitable assumptions are made for the shape of the electric field.
However, it seems to us that in the ac-regime this approach is not necessarily
appropriate since it may be very difficult to apply experimentally ac-fields
locally.

We find it more suitable to define a global average conductance via the time
average of the absorbed power. If we consider a monochromatic electric driving
field, this absorptive ac-conductance can be considered as the real part of a
complex conductance. It provides information about the magnitude of the
current in phase with the electric field.  The imaginary part of the
ac-conductance provides information about the phase shift between current and
voltage. It was obtained by a Kramers-Kronig transformation from the real
part.

We found that the absorptive ac-conductance factorizes into a product of the
density of excitations and the Fourier-transformed auto-correlation function
of the external electric field. A question related to this is whether the
electric field to be used is the external one or whether one has to use the
internal electric field that contains screening contributions. We found that
one can use the external electric field in order to obtain the ac-conductance
of the interacting system. If one uses the total field, the same result is
obtained for the conductance.  In any case, the result for $\omega\neq 0$
depends on the spatial shape of the field.

Only in the dc-limit, the conductance becomes independent of the shape
of the applied electric field and depends only on the applied external
voltage. It is renormalized by the interaction parameter, $\Gamma
_{1}=ge^{2}/h$. This does not contradict other recent results which
indicate that for a Luttinger system of a finite length connected to
Fermi liquid leads the conductance is not renormalized by the
interaction. We can argue that the interaction influences transport for
system lengths $L$ above $q^{-1}(\omega)$, the wave length of the
charge density wave. When $L<q^{-1}(\omega)$ the effect of the
interaction can be neglected \cite{kf92,mat93}. Since we are
considering the thermodynamic limit, this region is outside the range
of the validity of our model.

There is a resonance in the absorptive conductance at a frequency which
corresponds to the inverse of a characteristic length scale of the
interaction. In the limit of a 1D Coulomb interaction the latter is given by
the cutoff length $d$. For the Luttinger limit, it is the interaction range
$\alpha ^{-1}$. Its height and width 
is influenced by the electric field, but its
position is not. However, when having in mind a ``realistic'' situation, a
word of caution is here in order. If one wants to use the Luttinger model as a
model for a quantum wire, the parameters should be such that (i) linearization
of the free-electron dispersion is a good approximation, i.e. $q<k_{F}$, and
(ii) interband transitions are not important for the absorptive conductance,
$m^{*}\omega < hd^{-2}$ ($m^{*}$ effective mass). 
We have found above that $q_{\rm p}\propto d^{-1}$
in the Coulomb limit. The Fermi energy should be smaller than the interband
energy distance. Therefore, the Fermi wave number is restricted to
$k_{F}<d^{-1}$.  Then, $q_{\rm p}\approx k_{F}$, and corrections to
linearization should be taken into account. Since furthermore $\omega _{\rm
  m}\propto v_{F}/d\propto h/m^{*}d^{2}$ 
also interband transitions could become
important near the resonance. For the Luttinger limit of the interaction the
situation does not improve. One needs also to take into account interband
mixing induced by the interaction for a proper description of the frequency
region near the resonance. 

However, at low frequency, and if the Fermi level is well below the onset of
the second lowest subband, interaction-induced mixing of the bands can be
neglected, and interband transitions are unimportant. Here, the real and the
imaginary part of the conductance depend quadratically and linearly on
$\omega$, respectively. The signs of the prefactors of these terms indicate
capacitive- or inductive-like behavior of the system. By comparing the
frequency behavior of the microscopic model with that of a ``minimal''
classical circuit of capacitances, inductances and resistances we find
microscopic expressions for these quantities. They are, however, not unique.
Their validity is restricted to certain parameter regions. Only the scaling
properties are the same. Astonishingly, one can identify a parameter region,
in which the scaling properties with the parameters of the system are the same
for low frequency and near the resonance. This fact gives us some confidence
that although the frequency region near the resonance is somewhat out of the
range of validity of the model certain general features of the results remain
valid.

In particular, the impedances depend in general strongly on the shape
of the applied electric field. There is a competition between the range
of the field and the range of the interaction, which determines whether
the system behaves as a capacitor ($\ell \alpha \ll 1$) or as an
inductor ($\ell \alpha \gg 1$). In the former case, depending on the
ratio between the interaction parameter $g$ and $\ell\alpha$, the
capacitance and the dissipative resistance may ($g/\ell\alpha \ll 1$)
or may not ($g/\ell\alpha \gg 1$) depend on the field range. The
infinite range of the interaction removes the dependence on the range
of the field, such that one can interpret the impedances as genuine
properties of the system that are only determined by the parameters of
the microscopic model.

It is particularly interesting to note that the model allows to
identify two conceptually different ``resistances''. In the dc-limit,
we have the equivalent of the ``contact resistance'', $h/ge^{2}$, which
is however renormalized by the interaction parameter. Near the resonance
frequency, one can define a ``dissipative resistance'', approximately
given by the inverse of the height of the resonance. The former depends
on if and how the system is connected with the ``outside world'' via
contacts. It changes to $h/e^{2}$ if the system length is smaller than
the wave length of the charge density wave. In contrast, we expect that
the latter is not changed as long as $L>q_{\rm p}^{-1}\propto d$. By
comparing with the classical circuit one also notes that the
dissipative resistance is also present at low frequency, though it
scales differently with the system parameters.

\section{Conclusion}

How to calculate the frequency dependent low-temperature electronic quantum
transport properties of nanostructures when interaction effects are important
is presently a subject of strong debate. Since experiment can access this
regime by using modern fabrication technology, developing theoretical concepts
for the treatment of transport processes taking into account interactions is
important. In addition, future information technology will require
high-frequency signal transportation in nano-scale systems. Proper theoretical
understanding of the underlying microscopic processes will be essential for
developing this technology in the far future.

In order to obtain insight into some of the peculiar microscopic features of
this transport region, we have investigated the ac-response of a simple 1D
model of an interacting electron system.  We found that in spite of the
simplicity of the model the definition of transport parameters provides
already non-trivial questions to be solved. The transport parameters depend on
the purpose for which they are supposed to be used. For instance, when asking
for the current in some probe as a response to an electric field applied to
another probe it is the non-local conductivity, which has to be used as the
``conductance''. On the other hand, when being interested in the absorption of
electromagnetic radiation, an average of the conductivity provides an
``absorptive conductance''. 

By starting from the ``absorptive conductance'', we have made an attempt to
define for our system the quantum counterparts of the impedance parameters
used in the transport theory of classical circuits.  We found that they cannot
be defined uniquely in terms of the parameters of the microscopic model, in
accordance with the above general statement.  However, certain scaling
properties remain valid in astonishingly large regions of parameters. They
even resemble the scaling properties of the corresponding classical
definitions. This encourages to find similar quantities for more complicated
systems as tunnel barriers and multiple tunnel barriers in the presence of
interactions.

Useful discussions with Andrea Fechner are gratefully ack\-now\-led\-ged.
This work has been supported by Istituto
Nazionale di Fisica della Materia, by the
Deut\-sche For\-schungs\-ge\-mein\-schaft via
the Graduiertenkolleg ``Nanostrukturierte Festk\"orper'' and the SFB 508 of
the Universit\"at Hamburg, by the European Union within the HCM and
TMR-programmes, and by NATO.

\noindent
$^{1}$ on leave of absence from Dipartimento di Fisica, INFM, Universit\`a di
Genova, Via Dodecaneso 33, I--16146, Genova.\\
$^{2}$ on leave of absence from Istituto di Fisica di Ingegneria, INFM,
Universit\`a di Genova, Via Dodecaneso 33, I--16146, Genova.


\begin{thebibliography}{99}
\bibitem{al86} D. V. Averin, K. K. Likharev, J. Low Temp. Phys. {\bf 62},
345 (1986).
\bibitem{fd87} T. A. Fulton, G. J. Dolan, Phys. Rev. Lett. {\bf 59}, 109 (1987)
\bibitem{getal89} L. J. Geerligs, V. F. Anderegg, C. A. van der Jeugd,
J. Romin, J. E. Mooij, Europhys. Lett.
{\bf 10}, 79 (1989).
\bibitem{in92} G. L. Ingold, Yu. V. Nazarov, in {\em Single Charge
Tunneling.}, Volume 294 of
NATO Advanced Study Institute Series B, edited by H. Grabert, M. H.
Devoret,  21 (Plenum, New York 1992).
\bibitem{meirav90} U. Meirav, M. A. Kastner, S. J. Wind, Phys. Rev. Lett.
{\bf 65}, 771 (1990).
\bibitem{jetal92} A. T. Johnson, L. P. Kouwenhoven, W. de Jong, N. C.
van der Vaart, C. J. P. M. Harmans C. T. Foxon, Phys. Rev. Lett. {\bf
69}, 1592 (1992).
\bibitem{wetal93} J. Weis, R. J. Haug, K. von Klitzing, K. Ploog,
Phys. Rev. B{\bf 46}, 12837
(1992); Phys. Rev. Lett. {\bf 71}, 4019 (1993).
\bibitem{weinprl} D. Weinmann, W. H\"ausler, B. Kramer, Phys. Rev. Lett.
{\bf 74}, 984 (1995).
\bibitem{weinmann96} D. Weinmann, W. H\"ausler, B. Kramer,
Ann. Phys. (Leipzig) {\bf 5}, 652 (1996).
\bibitem{sollner83} T. C. L. G. Sollner, W. D. Goodhue, P. E. Tannewald,
C. D. Parker, D. D. Peck, Appl. Phys. Lett.
{\bf 43}, 588 (1983).
\bibitem{chitta92} V. A. Chitta, R. E. M. de Bekker, J. C. Mann, S. J.
Hawksworth, J. M. Chamberlain, M. Henni, G. Hill, Surf.
Sci. {\bf 263}, 227 (1992).
\bibitem{cai90} W. Cai, T. F. Zheng, P. Fu, M. Lax, K. Shum, R. R.
Alfano, Phys. Rev. Lett. {\bf 65}, 104 (1990).
\bibitem{wagner} M. Wagner, Phys. Rev. Lett. {\bf 76}, 4010 (1996). 
\bibitem{kouwen94} L. P. Kouwenhoven, S. Jauhar, K. McCormick, D. Dixon,
P. L. McEuen, Yu V. Nazarov, N. C. van der Vaart, Phys. Rev. B{\bf 50},
2019 (1994); L. P. Kouwenhoven, S. Jauhar, J. Orenstein, P. L. McEuen,
Y. Nagamune, J. Motohisa, H. Sakaki, Phys. Rev. Lett. {\bf 73},
3443 (1994).
\bibitem{blick95} R. H. Blick, R. J. Haug, D. W. van der Weide, K. von
Klitzing, K. Eberl, Appl. Phys. Lett. {\bf 67},
3924 (1995).
\bibitem{sm89} C. Sikorski, U. Merkt, Phys. Rev. Lett. {\bf 62}, 2164 (1989).
\bibitem{h91} D. Heitmann K. Ensslin, in {\em Quantum Coherence in Mesoscopic
Systems}, Volume 254 of NATO Advanced Study Institute Series B, edited
by B. Kramer p. 3, (Plenum, New York 1991).
\bibitem{mhp92} B. Meurer, D. Heitmann, K. Ploog, Phys. Rev. Lett. {\bf
68}, 1371 (1992).
\bibitem{goni91} A. R. Go\~{n}i, A. Pinczuk, J. S. Weiner, J. M. Calleja,
B. S. Dennis, L. N. Pfeifer, K. W. West, Phys. Rev. Lett. {\bf 67},
3298 (1991).
\bibitem{setal94} R. Strenz, U. Bockelmann, F. Hirler, G. Abstreiter, G.
B\"ohm, G. Weimann, Phys. Rev. Lett. {\bf 73}, 3022  (1994)
\bibitem{schetal96} C. Sch\"uller, G. Biese, K. Keller, C. Steinebach,
D. Heitmann, Phys. Rev. B{\bf 54},
R17304 (1996).
\bibitem{cpt84} G. Carr, S. Percowitz N. Tanner, in {\em Far Infrared
Properties of Inhomogeneous Materials}, edited by K. Button (Academic
press, New York, 1991).
\bibitem{zsta96} F. Zhou, B. Spivak, N. Taniguchi, B. L. Altshuler,
Phys. Rev. Lett. {\bf 77}, 1958 (1996).
\bibitem{rb96} Y. Noat, B. Reulet, H. Bouchiat, D. Mailly, in: {\em
Correlated Fermions and Transport in Mesoscopic Systems}, edited by T.
Martin, G. Montambaux, J. Tr\^{a}n Thanh V\^{a}n, 141 (Editions
Frontieres, Gif-sur-Yvette 1996); B. Reulet, M. Ramin, H. Bouchiat, D.
Mailly, Phys. Rev. Lett. {\bf 75}, 124 (1996).
\bibitem{i86} Y. Imry, in: {\em Directions in Condensed Matter Physics},
edited by G. Grinstein and G. Mazenko, 221 (World Scientific, Singapore 1986).
\bibitem{velicky89} B. Velicky, J. Ma\v{s}ek, B. Kramer, Phys. Letters
A{\bf 140}, 447 (1989).
\bibitem{liu91} H. C. Liu, Phys. Rev. B{\bf 43}, 12538 (1991).
\bibitem{brandes93} T. Brandes, D. Weinmann and B. Kramer, Europhys. Lett.
{\bf 22}, 51 (1993).
\bibitem{fu93} Y. Fu and C. Dudley, Phys. Rev. Lett. {\bf 70}, 65 (1993). 
\bibitem{wjm93} N. S. Wingreen, A.-P. Jauho, Y. Meir, Phys. Rev. B{\bf
48}, 8487 (1993-I).
\bibitem{buettiker93} M. B\"uttiker, M. Pr\^{e}tre, H. Thomas, Phys.
Rev. Lett. {\bf 70}, 4114 (1992).
\bibitem{bc96} M. B\"uttiker, T. Christen, in {\em Quantum Transport in
Semi\-con\-ductor Sub\-micron Struc\-tures}
edited by B. Kramer NATO Advan\-ced
Study Insti\-tute Series E {\bf 326}, 263 (Kluwer, Dordrecht 1996);
Phys. Rev. Lett. {\bf 77}, 143 (1996).
\bibitem{tien63} P. K. Tien, J. R. Gordon, Phys. Rev. {\bf 129}, 647 (1963).
\bibitem{inarrea94} R. Aguado, J. I\~{n}arrea, G. Platero, Phys. Rev.
B{\bf 53}, 10030 (1996).
\bibitem{bruder94} C. Bruder, H. Schoeller, Phys. Rev. Lett. {\bf 72},
1076 (1994).
\bibitem{hettler95} M. H. Hettler, H. Schoeller,  Phys. Rev. Lett.
{\bf 74}, 4907 (1995).
\bibitem{sw96} C. A. Stafford, N. S. Wingreen, Phys. Rev. Lett. {\bf
76}, 1916 (1996).
\bibitem{sasswk96} M. Sassetti, U. Weiss, B. Kramer, Sol. St. Comm.
{\bf 97}, 605 (1996).
\bibitem{kane92} C. L. Kane, M. P. A. Fisher, Phys. Rev. B{\bf 46},
15233 (1992-I).
\bibitem{yue94} D. Yue, L. I. Glazmann, K. A. Matveev, Phys. Rev. B{\bf
49}, 1966 (1994).
\bibitem{guinea94} F. Guinea, G. G\'{o}mez-Santos, M. Sassetti, U. Ueda,
Europhys. Lett. {\bf 30}, 561 (1995).
\bibitem{sassetti96} M. Sassetti, B. Kramer, Phys. Rev. B{\bf 54}, R5203
(1996). 
\bibitem{keller96} O. Keller, Phys. Rep. {\bf 268}, 85 (1996).
\bibitem{kawabata} A. Kawabata, J. Phys. Soc. Japan {\bf 65}, 30 (1996).
\bibitem{finkelstein} Y. Oreg, A. M. Finkel'stein, Phys. Rev. B{\bf 54},
R14265 (1996).
\bibitem{ponomarenko} V. V. Ponomarenko, Phys. Rev. B{\bf 52}, R8666 (1995).
\bibitem{fn96} A. Furusaki, N. Nagaosa, Phys. Rev. B{\bf 54}, R5239 (1996).
\bibitem{maslov} D. Maslov, M. Stone, Phys. Rev. B{\bf 52}, R5539
(1995); D. Maslov, Phys. Rev. B{\bf 52}, R14368 (1995).
\bibitem{cuni96} G. Cuniberti, M. Sassetti, B. Kramer, J. Phys.:
Condens. Matter {\bf 8}, L21 (1996).
\bibitem{fisherandlee} D. S. Fisher, P. A. Lee, Phys. Rev.
B{\bf 23}, 6851 (1981). 
\bibitem{landauer70} R. Landauer, Phil. Mag. {\bf 21}, 863 (1970).
\bibitem{sassk96} M. Sassetti, B. Kramer, Phys. Rev. B{\bf 55}, 9306 (1997).
\bibitem{luttinger63} J. M. Luttinger, J. Math. Phys. {\bf 4}, 1154 (1963).
\bibitem{lp74} A. Luther, I. Peschel, Phys. Rev. B{\bf 9}, 2911 (1974).
\bibitem{s79} J. S\'{o}lyom, Adv. Phys. {\bf 28}, 209 (1979).
\bibitem{haldane81} F. D. M. Haldane, J. Phys. C{\bf 14}, 2585 (1981).
\bibitem{fabrizio} The case of open boundary conditions was treated in
M. Fabrizio, A. O. Gogolin, Phys. Rev. B{\bf 51}, 17827 (1995);
S. Eggert, H. Johannesson, A. Mattson, Phys. Rev. Lett. {\bf 76}, 1505 (1996).
\bibitem{schulz93} H. J. Schulz, Phys. Rev. Lett. {\bf 71}, 1864 (1993).
\bibitem{abra72} N. Abramowitz, I. A. Stegun, {\em Handbook of
Mathematical Functions} (9th Edition), p. 228, Eq.~(5.1.1) (Dover Publ. New
York, 1972).    
\bibitem{sarma92} Q. P. Li, S. Das Sarma, R. Joynt, Phys. Rev. B{\bf
45}, 13713 (1992-I); Q. Li, S. Das Sarma, Phys. Rev. B{\bf 40}, 5860 (1989).
\bibitem{kf92} C. L. Kane, M. P. A. Fisher,  Phys. Rev. Lett. {\bf 68}, 1220
(1992).
\bibitem{mat93} K. A. Matveev, L. I. Glazman, Physica B{\bf 189}, 266 (1993);
L. I. Glazman, I. M. Ruzin, B. I. Shklovskii, Phys. Rev. B{\bf 45},
8454 (1992).
\end{thebibliography}
\end{document}